\documentclass[iop]{emulateapj}
\usepackage{epsfig,hyperref,color}

\submitted{}
\received{2014 April 7}
\accepted{2014 June 11}
\published{2014 June 20}

\begin{document}

\title{Fingerprints of Galactic Loop~I on the Cosmic Microwave Background}

\author{Hao Liu\altaffilmark{1,2}, Philipp Mertsch\altaffilmark{3}, Subir Sarkar\altaffilmark{4,5}}
\altaffiltext{1}{Niels Bohr Institute and DISCOVERY Center, Copenhagen
  University, Blegdamsvej 17, 2100 Copenhagen \O, Denmark}
\altaffiltext{2}{Institute of High Energy Physics, CAS, Beijing, China}
\altaffiltext{3}{Kavli Institute for Particle Astrophysics \& Cosmology,
  2575 Sand Hill Road, M/S 29, Menlo Park, CA 94025, USA}
\altaffiltext{4}{Rudolf Peierls Centre for Theoretical Physics, University
  of Oxford, 1 Keble Road, Oxford OX1 3NP, UK}
\altaffiltext{5}{Niels Bohr International Academy, Copenhagen University,
  Blegdamsvej 17, 2100 Copenhagen \O, Denmark}

\begin{abstract}
We investigate possible imprints of galactic foreground structures
such as the ``radio loops'' in the derived maps of the cosmic microwave
background. Surprisingly there is evidence for these not only at radio
frequencies through their synchrotron radiation, but also at microwave
frequencies where emission by dust dominates. This suggests the
mechanism is magnetic dipole radiation from dust grains enriched by
metallic iron or ferrimagnetic materials. This new foreground we have
identified is present at high galactic latitudes, and potentially
dominates over the expected $B$-mode polarization signal due to
primordial gravitational waves from inflation.
\end{abstract}


\maketitle

\section{Introduction}
\label{sec:intro}

The study of the cosmic microwave background (CMB) radiation is a key
testing ground for cosmology and fundamental physics, wherein
theoretical predictions can be confronted with observations
\citep{WMAP7:powerspectra,PlanckXV,PlanckXVI}. The temperature
fluctuations in the CMB have provided our deepest probe of the Big
Bang model and of the nature of space--time itself. Moreover CMB data
provide a bridge between cosmology and astro-particle physics,
shedding light on galactic cosmic rays \citep{Mertsch}, and galactic
X- and $\gamma$-ray emission \citep{PlanckXXVI, PlanckI,PlanckIX}.

Since the first release of data from the \textit{Wilkinson
  Microwave Anisotropy Probe} (WMAP), it has been noted that the
derived CMB sky maps exhibit departures from statistical isotropy
\citep{Chiang_NG,Chiang_Nas_Coles,Tegmark:Alignment,Multipole_Vector1,
  Multipole_Vector4,Multipole_Vector2,Park_Genus,
  Hemispherical_asymmetry,Axis_Evil,Axis_Evil2,
  cold_spot_origin,power_asymmetry_wmap5,power_asymmetry_subdegree,odd,Gruppuso:2013xba},
probably because of residuals from the incomplete removal of galactic
foreground emission. For example the Kuiper Belt in the outer Solar
System may partly be responsible for the unexpected
quadrupole-octupole alignment and parity asymmetry in the CMB
\citep{Hansen}. This issue has acquired even more importance after the
first release of cosmological data products from the \textit{Planck}
satellite \citep{PlanckXXIII}.

In this paper we construct a physical model to account for the local
features of the WMAP internal linear combination (ILC) map
\citep{Bennett2012} in the direction of the galactic ``radio loops'', in
particular Loop~I \citep{Berkhuijsen1971}. We show that in the low
multipole domain, $\ell \le 20$, the peaks of the CMB map
\emph{correlate} with radio synchrotron emission from Loop~I. However,
the physical source of this anomaly is likely related to emission by
dust---including magnetic dipole emission from dust grains enriched
by metallic Fe, or ferrimagnetic materials like
$\mathrm{Fe}_3\mathrm{O}_4$ \citep {DL_99,2013ApJ...765..159D}.

The radio loops are probably old supernova remnants (SNRs) in the
radiative stage which have swept up interstellar gas, dust, and
magnetic field into a shell-like structure. Dust grains are
well-coupled to the gas by direct collisions, Coulomb drag, and
Lorentz forces. Hence along with synchrotron radiation by high energy
electrons, dust emission will be enhanced in the shell, with
limb-brightening yielding the observed ring-like morphology.
Moreover, part of the Loop~I anomaly overlaps with the \textit{Planck}
haze at 30 GHz~\citep{PlanckIX}. We are investigating these issues in
more detail and will present the results elsewhere.

The structure of this Letter is as follows: In Section~\ref{sec:loops} we
summarise the properties of Loops I-IV from radio waves to
$\gamma$-rays. In Section~\ref{sec:peaks} we demonstrate that there are
spatial correlations between features in the WMAP 9~yr ILC map of
the CMB and Loop~I---the CMB temperature is systematically shifted
by $\sim 20\,\mu$K along Loop~I. We perform a cluster analysis along
the Loop~I radio ring and show that the peaks in the CMB map are
indeed clustered in this ring with a chance probability of
$\sim10^{-4}$. In Section~\ref{sec:residual}, we discuss how this signal
from Loop~I could have evaded foreground subtraction using the ILC
method. In Section~\ref{sec:models} we discuss physical mechanisms of the
emission from the Loop~I region and argue that it likely arises from
interstellar dust, possibly including magnetic dipole radiation from
magnetic grain materials. We present our conclusions in
Section~\ref{sec:conclusion}.

\section{The radio loops in the Milky Way}
\label{sec:loops}

\begin{figure*}[!tb]
\centering
\includegraphics[width=\columnwidth]{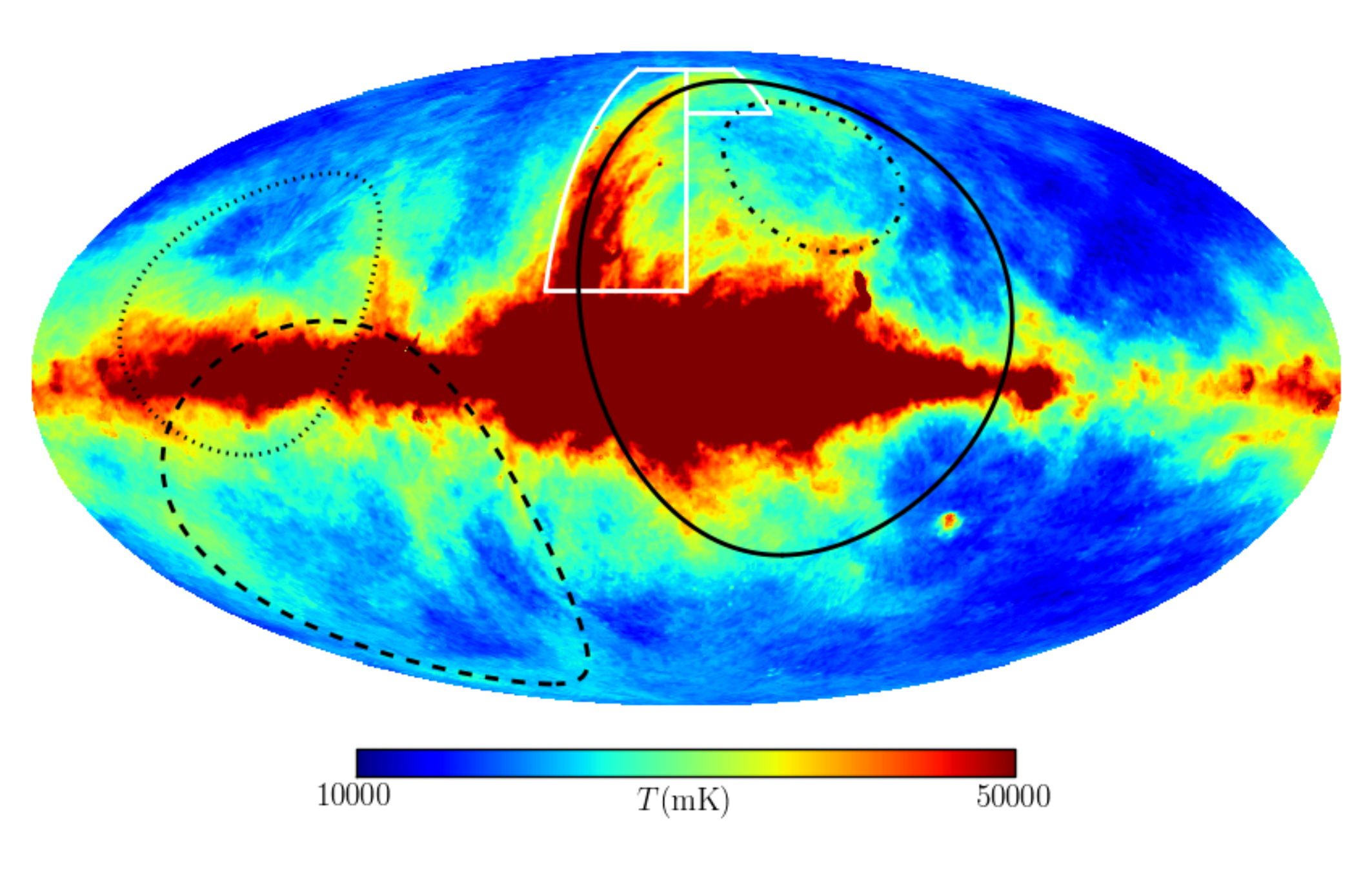}
\includegraphics[width=\columnwidth]{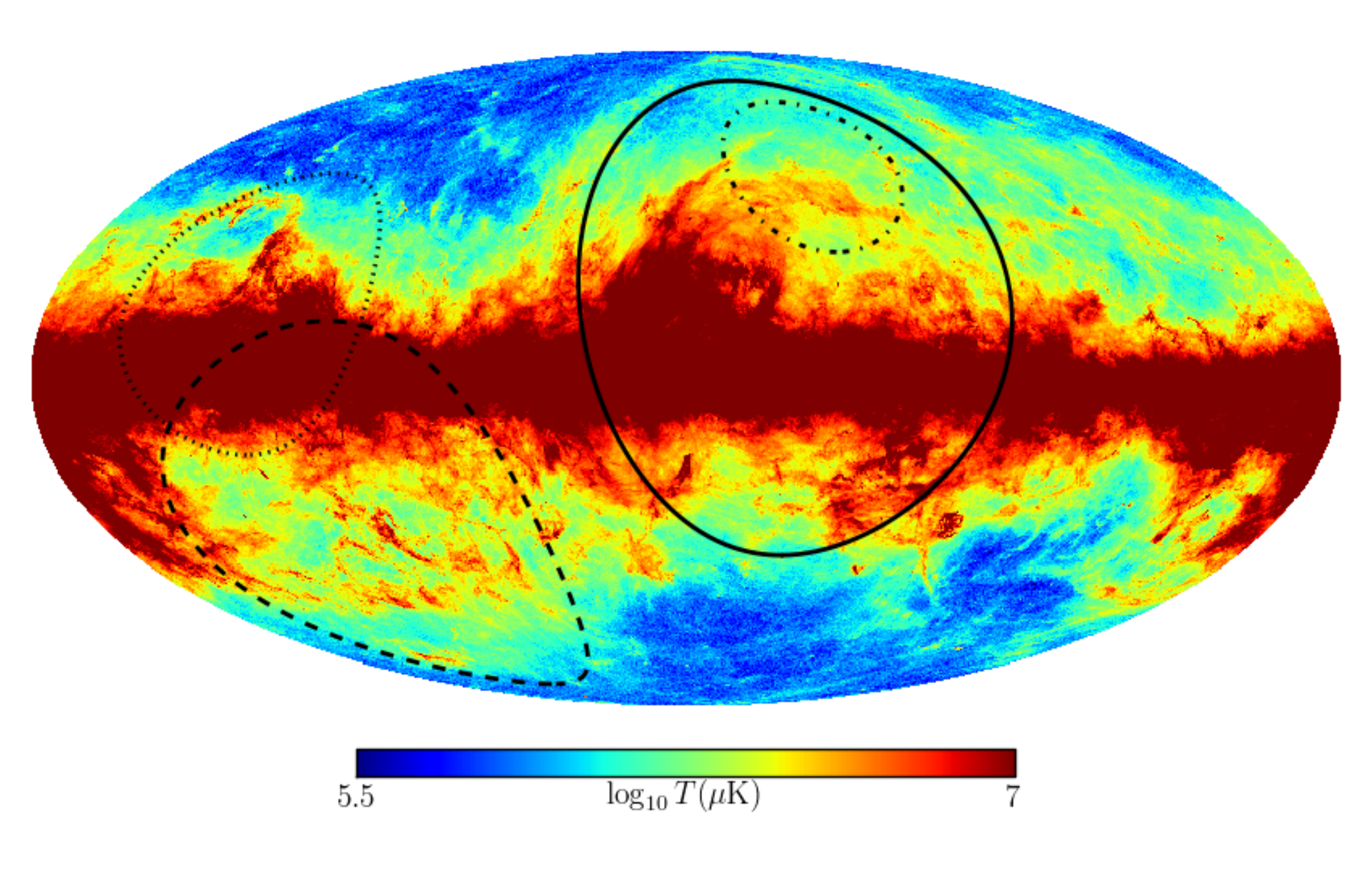}
\includegraphics[width=\columnwidth]{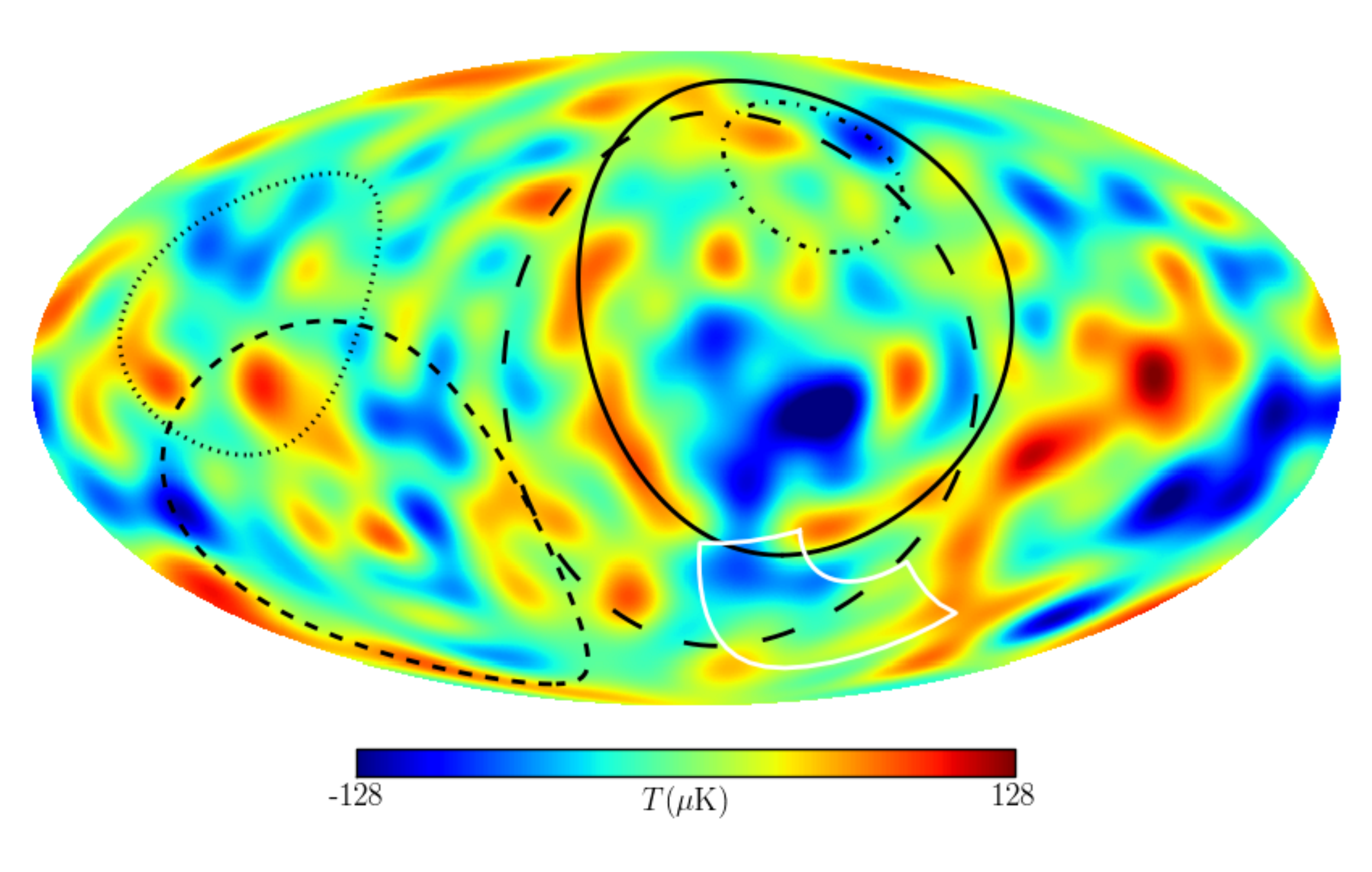}
\includegraphics[width=\columnwidth]{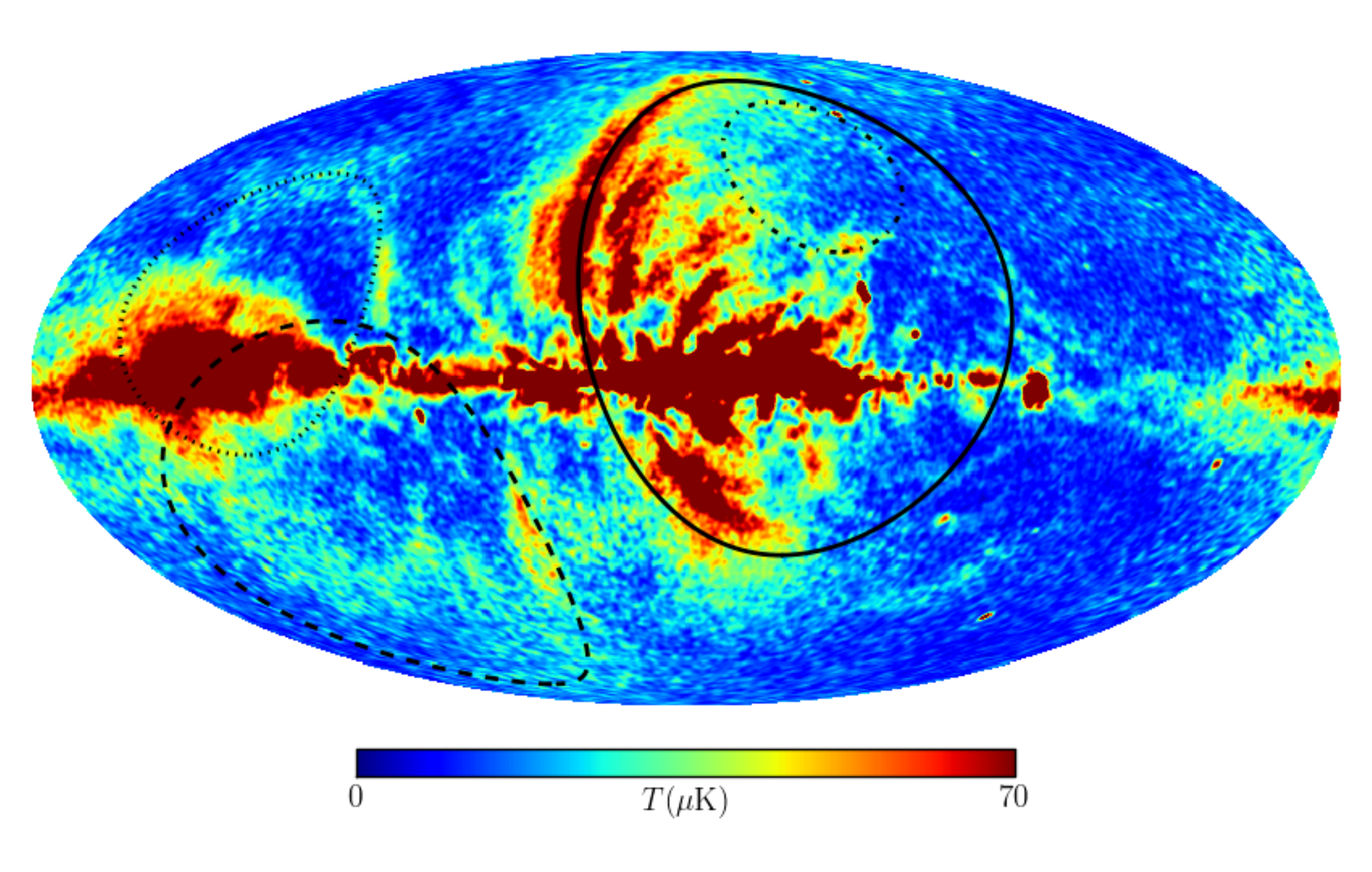}
\caption{The 408 MHz survey (top left), the \textit{Planck} 857 GHz
  map (top right) the low resolution ($l \le 20$) WMAP9 ILC map
  (bottom left) and the WMAP9 K-band polarised intensity map (bottom
  right), with the positions of the radio loops indicated: Loops~I-IV
  are indicated by the solid, dashed, dotted and dash-dotted lines,
  respectively and ``Wolleben's loop S1'' by the long-dashed line. The
  white outline (top left panel) marks the NPS at 22 MHz, and the
  BICEP2 observation region is also indicated (bottom left panel).}
\label{fig1}
\end{figure*}

Radio surveys of the Galaxy reveal a number of features that are parts
of larger ``radio loops'' \citep{Berkhuijsen1971}. The most prominent is
the North-Polar Spur (NPS) which is part of Loop~I
\citep{Roger1999}. It has been noted that the radio loops correlate
with expanding shells of gas and dust, energised by supernovae or stellar
winds \citep{Berkhuijsen1971,1980ApJ...242..533H, Salter1983,
  Wolleben2007}. The Loop~I superbubble has been attributed to stellar
winds from the Sco-Cen OB association and SN activity, with the NPS
being the brightest segment of a SNR. The ambient magnetic field is
most likely draped around the expanding bubbles
\citep{1980ApJ...242..533H}, as seen in radio and optical polarization
data. The NPS is observed over a wide range of wavelengths---the 408
MHz and 1.4 GHz all-sky surveys, the ROSAT X-ray surveys at 0.25, 0.75
and 1.5~keV, and soft and hard $\gamma$-ray sky maps from EGRET and
\textit{Fermi}-LAT. It has also been detected in the WMAP K-band
intensity and polarization maps, and more recently in the 2013
\textit{Planck} 30 GHZ temperature map \citep{PlanckXV}, and even in
the 353-857 GHz far-IR maps. \citet{2012MNRAS.426...57H} have
suggested from a cross-correlation of Faraday rotation and WMAP maps
that such structures may affect the measured CMB temperature at high
galactic latitudes.

We adopt the properties of Loops~I-IV as given in Table~1 of
\citet{Berkhuijsen1971}.  In Figure~\ref{fig1} we show the 408 MHz
all-sky survey \citep{Haslam1982}, the 2013 \textit{Planck} 857 GHz
map, the WMAP9 K-band polarization intensity map, and the WMAP9 low
resolution ($\ell\le 20$) ILC map of the CMB --- hereafter called
``ILC9''. We have superimposed the loops and the NPS, as well as the sky
region observed by BICEP2 \citep{Ade:2014xna}.

\section{CMB peaks along Loop~I}
\label{sec:peaks}

To investigate possible correlations between the radio loops and
features in the ILC9 map, we examine a ring of width $\pm 2^\circ$
along Loop~I. The average temperature of the ILC9 signal along the
$\pm 2^\circ$ ring, $\overline{T} = (2\pi)^{-1} \int_0^{2\pi}
\mathrm{d}\Phi\,(T_{\text{ILC}}(\Phi) - \overline{T}_{\text{ILC}})
\simeq 23.9 \,\mu\text{K}$, deviates significantly from the
expectation for a random realization of the CMB. In order to quantify
this, we have generated 1000 CMB realizations of the WMAP best-fit
$\Lambda\text{CDM}$ cosmological model \citep{Bennett2012}.  Computing
the number of simulated realizations with an average temperature equal
or larger than the observed $\overline{T} = 23.9 \, \mu\text{K}$, we
find a $p$-value of only 0.01.

We have also determined the skewness of the distribution of ILC9
temperatures along the $\pm 2^{\circ}$ Loop~I ring: $\gamma_1 = 1/(2
\pi) \int_0^{2\pi} \mathrm{d}\Phi \, \left[(T_{\text{ILC}}(\Phi) -
  \overline{T}_{\text{ILC}}(\Phi))/\sigma \right]^3 \simeq -0.68$,
where $\sigma$ is the standard deviation of the ILC9
temperatures. Comparing this with the skewness of 1000 Monte Carlo (MC)
simulations we find a similarly small $p$-value of 0.03.

These anomalies warrant a more detailed analysis. In order to show the
result more clearly, we have investigated the statistical isotropy in
a $\pm 10^{\circ}$ ring around Loop~I, using a cluster analysis
\citep{Naselsky1995} that has been applied previously to WMAP data
\citep{Doroshkevich2004}: If the ILC9 map is statistically isotropic,
there should be \emph{no} correlation between the position of its
peaks and the position of Loop~I. We therefore test the hypothesis
that the distribution of the positive peaks around Loop~I is
\emph{not} random.

\begin{figure}[!tbh]
\centerline{\includegraphics[scale=0.4]{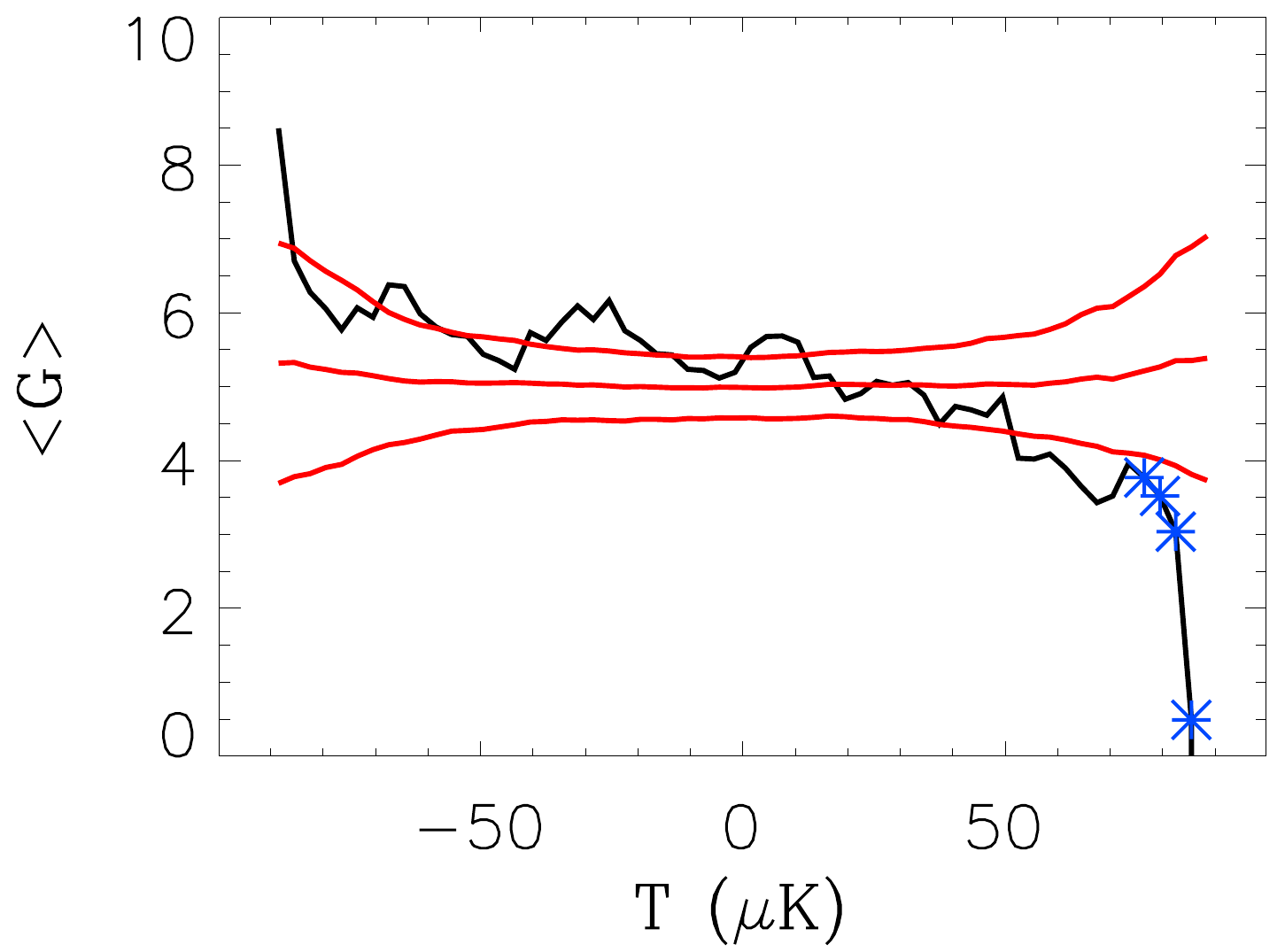}}
\caption{Average angular distance to Loop~I vs. temperature.
  The curve is descending, indicating that higher temperature pixels
  are closer to Loop~I. The red lines show the $\pm1\sigma$ range,
  while the 4 highest temperature bins are marked by asterisks.}
\label{fig2}
\end{figure}

\begin{figure}[!tbh]
\centerline{\includegraphics[width=0.5\textwidth,angle=0]{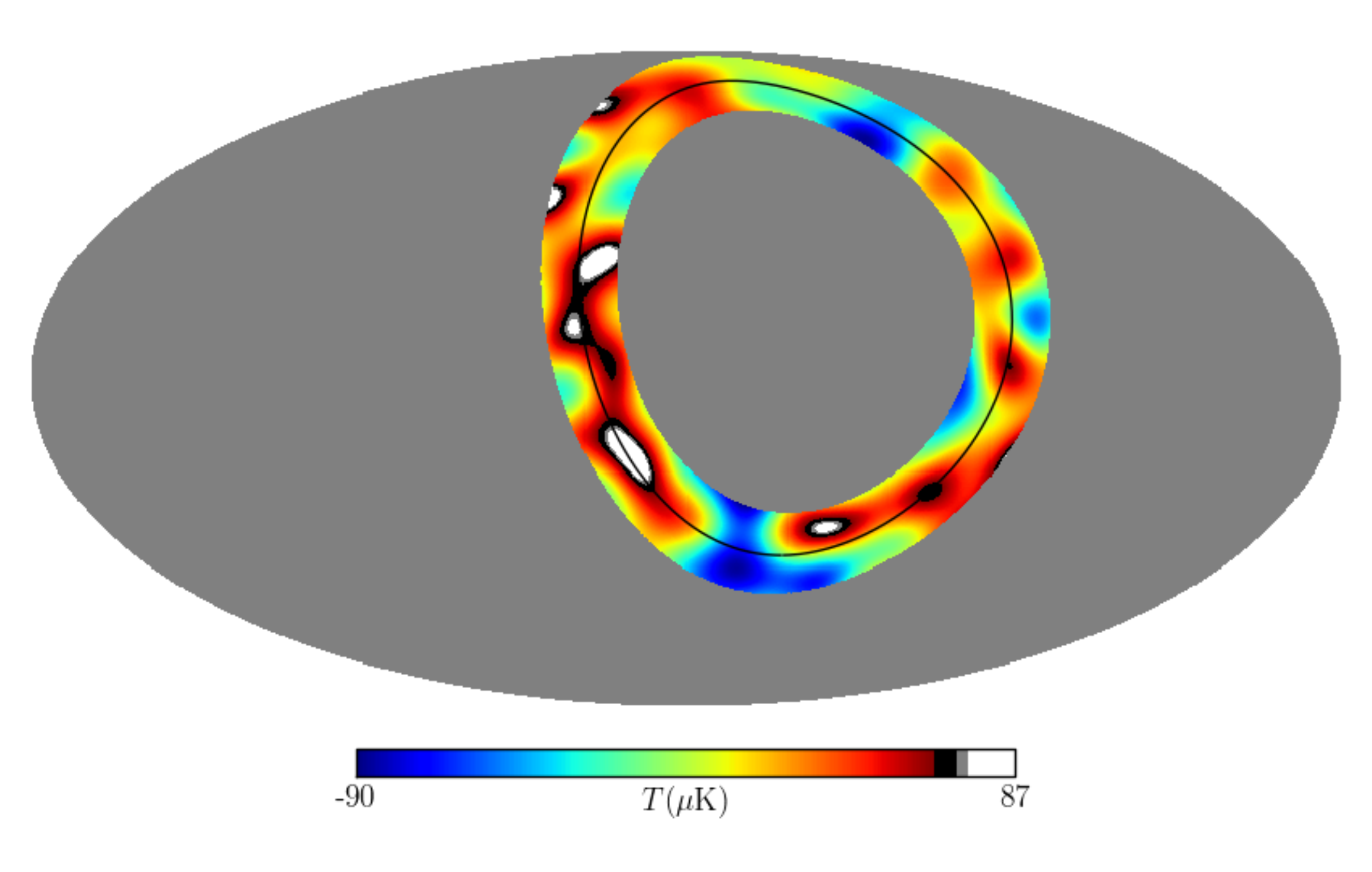}}
\caption{ILC9 temperature in a $\pm 10^\circ$ ring around
  Loop~I. The white zones correspond to the four bins with the highest
  temperature, the gray zones show the effect of adding the next bin,
  and the black zones indicate the effect of adding two more bins.}
 \label{fig3}
\end{figure}

To quantify this, we compute for each pixel the angular distance $G$
from Loop~I along great circles crossing both the pixel and the loop
center.  Figure~\ref{fig2} shows the average distance $\langle G \rangle$
for all pixels in a $\pm 10^{\circ}$ wide ring around Loop~I, averaged
in $\Delta T = 3 \, \mu\text{K}$ bins. It is apparent that on average,
pixels with higher temperature are closer to Loop~I, i.e., the peaks
in the ILC9 map \emph{are} clustered along Loop~I.  To quantify the
chance probability of this, we have generated 100,000 CMB maps from
the WMAP best fit $\Lambda\text{CDM}$ cosmological
model~\citep{Bennett2012}. We determine the fraction of mock maps that
have an average angular distance $g_i$ smaller than the observed
angular distance $G_i$ in bin $i$. Only 18 of the 100,000 mock maps
satisfy $g_i < G_i$ for all four last bins. In
Table~\ref{tbl:clustering}, we list the probabilities for the clustering
around Loop~I to occur by chance, adopting different criteria. In the
sky map in Figure~\ref{fig3}, the pixels in the four highest temperatures
bins are marked white, while the pixels in the fifth (sixth and
seventh) highest temperature bins are marked gray (black). We have
checked the effect of applying the WMAP ILC mask on the anomalies in
mean, skewness and clustering and we find our results to be largely
unchanged.

\begin{table}
\caption{The Probabilities for $g_i < G_i$ $(i=1\sim4)$ for the
  Highest Four Bins Under Different Criteria, Evaluated with 100,000 MC
  Simulations}
\label{tbl:clustering}
\centering
\begin{tabular*}{0.8\columnwidth}{@{\extracolsep{\fill} } l l}
\hline \hline
Criterion       &  Probability \\
& (\%) \\
\hline
$g_i<G_i$ for all four bins  & 0.018 \\
$g_i<G_i$ for any three in four bins & 2.0 \\
$g_i<G_i$ for any two in four bins & 3.3 \\
$g_i<G_i$, $i=1$ ($T=84-87 \, \mu\rm{K}$)& 0.1 \\
$g_i<G_i$, $i=2$ ($T=81-84 \, \mu\rm{K}$)& 2.9 \\
$g_i<G_i$, $i=3$ ($T=78-81 \, \mu\rm{K}$)& 7.5 \\
$g_i<G_i$, $i=4$ ($T=75-78 \, \mu\rm{K}$)& 8.6 \\
$\sum_i{g_i}<\sum_i{G_i}$, $i=1\sim4$ & 1.0\\
  \hline
\end{tabular*}
\end{table}

\section{The Loop~I anomaly as a residual from the CMB foregrounds separation}
\label{sec:residual}

We turn now to the question of how a \emph{physical} loop emission
could have evaded the foreground cleaning process. The ILC method for
isolation of the CMB signal is based on the independence of its
temperature on frequency. The temperature $S_j(\hat{\mathbf n})$ in
frequency band $j$ and direction $\hat{\mathbf{n}}$ is a superposition
of the primordial CMB $c(\hat{\mathbf n})$ and other components:
\begin{eqnarray}
S_j (\hat{\mathbf n}) = c(\hat{\mathbf n}) +F_j (\hat{\mathbf n}) \, ,
\label{eqn:Sj}
\end{eqnarray}
where $F_j(\hat{\mathbf n})=F_j^\mathrm{dust}(\hat{\mathbf n}) +
F_j^\mathrm{synch}(\hat{\mathbf n}) + \cdots + N(\hat{\mathbf n})$ is
the sum of dust, synchrotron and other foregrounds, as well as
instrumental noise.

For regions of the sky outside the galactic mask, the estimator of the
CMB anisotropy $d(\hat{\mathbf n})$ is given by an ILC of the $N$ different frequency bands:
\begin{eqnarray}
 d(\hat{\mathbf n}) = \sum_j W_j S_j (\hat{\mathbf n}) = c(\hat{\mathbf
   n}) + \sum_j W_j F_j(\hat{\mathbf n}) \,.
\label{dip2}
\end{eqnarray}
The weights are chosen so as to preserve unit response to the CMB,
i.e., $\sum_j W_j = 1$, and to minimise contamination by
foregrounds. The reconstruction of the CMB $c(\hat{\mathbf n})$
through the estimator $d(\hat{\mathbf n})$ is exact, if and only if
\begin{eqnarray}
\varepsilon(\hat{\mathbf n})\equiv d(\hat{\mathbf n}) - c(\hat{\mathbf
  n}) = \sum_j W_j F_j (\hat{\mathbf n})=0.
\label{dip3}
\end{eqnarray}
Since the coefficients $W_j$ are independent of $\hat{\mathbf{n}}$,
Equation~(\ref{dip3}) presumes that the frequency dependence of the
foregrounds is also \emph{independent} of $\hat{\mathbf{n}}$. For
example for a component with an assumed power-law frequency dependence
like the dust emission:
\begin{eqnarray}
 F_j^\mathrm{dust}(\hat{\mathbf n})=F_0^\mathrm{dust}(\hat{\mathbf
   n})\left(\frac{\nu_j}{\nu_0}\right)^{\alpha_j} \, ,
\label{for}
\end{eqnarray}
the $\alpha_j$ are taken to be constants, i.e., the morphology of the
foregrounds is assumed to be the \emph{same} in all bands. Here,
$\nu_j$ is the frequency of the $j$-th band, while
$F_0^\mathrm{dust}(\hat{\mathbf n})$ is the foreground sky map at the
reference frequency $\nu_0$.

In practice this assumption cannot be fully valid since the
transition from low frequencies
to high frequencies
leads to significant variability of the spectral index
$\alpha_j(\hat{\mathbf n})$. Clearly no method for optimization of the
difference $d(\hat{\mathbf n})-c(\hat{\mathbf n})$ can be free from
the residuals of foreground emissions for which
$\varepsilon(\hat{\mathbf n})\neq 0$. There is also an ultimate limit
to the accuracy with which the CMB map can be reconstructed due to its
chance correlations with the foregrounds. Consequently, some areas of
the reconstructed CMB map may be contaminated by foreground
residuals. An example is the contamination of the ILC map by residuals
after removal of point sources, which produces non-Gaussian features
in the derived CMB signal. We now show that a similar effect may be
causing the anomaly in the Loop~I region.

\begin{figure*}[!tb]
\centerline{
\includegraphics[width=0.33\textwidth]{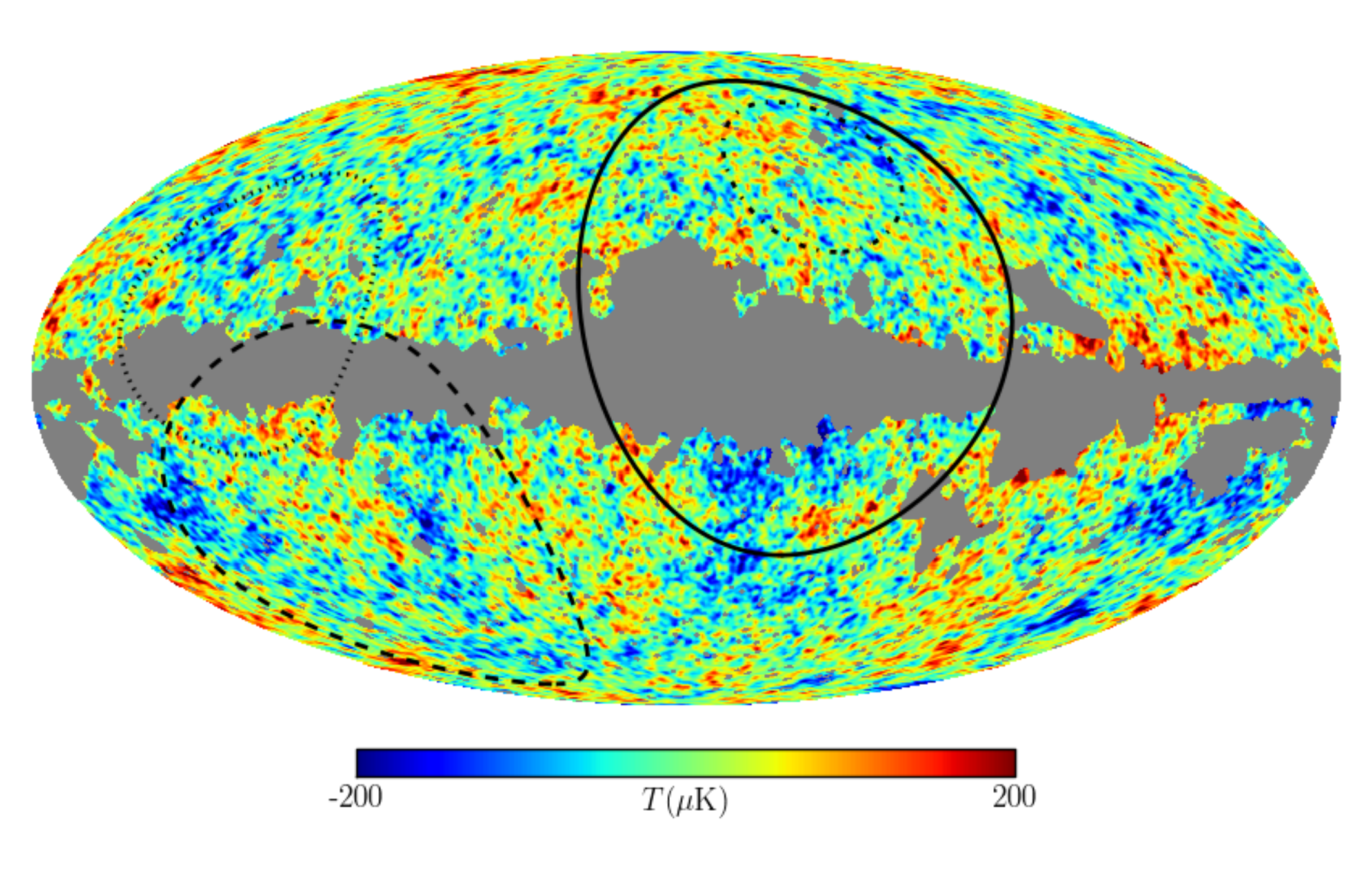}\\
\includegraphics[width=0.33\textwidth]{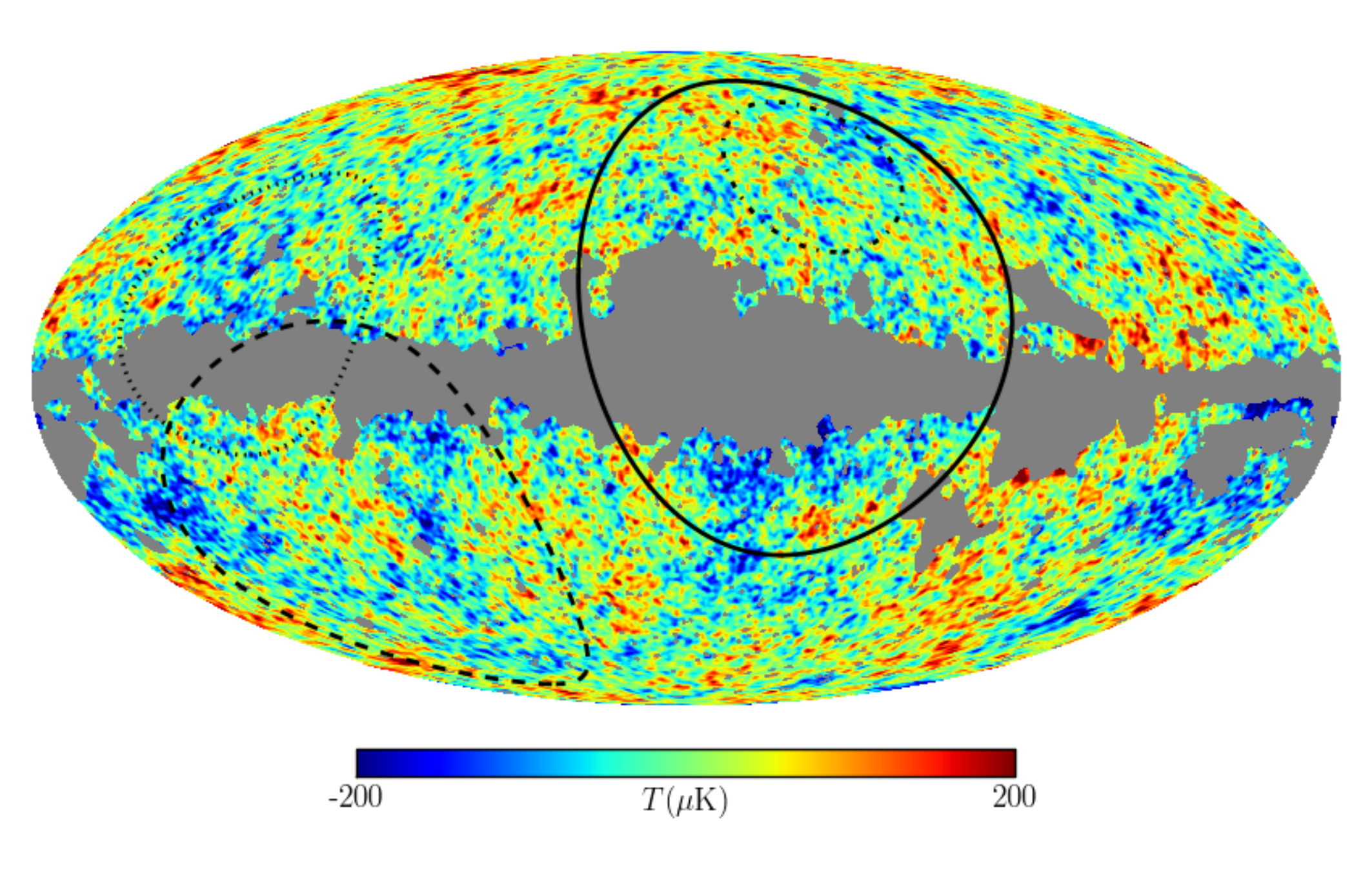}\\
\includegraphics[width=0.33\textwidth]{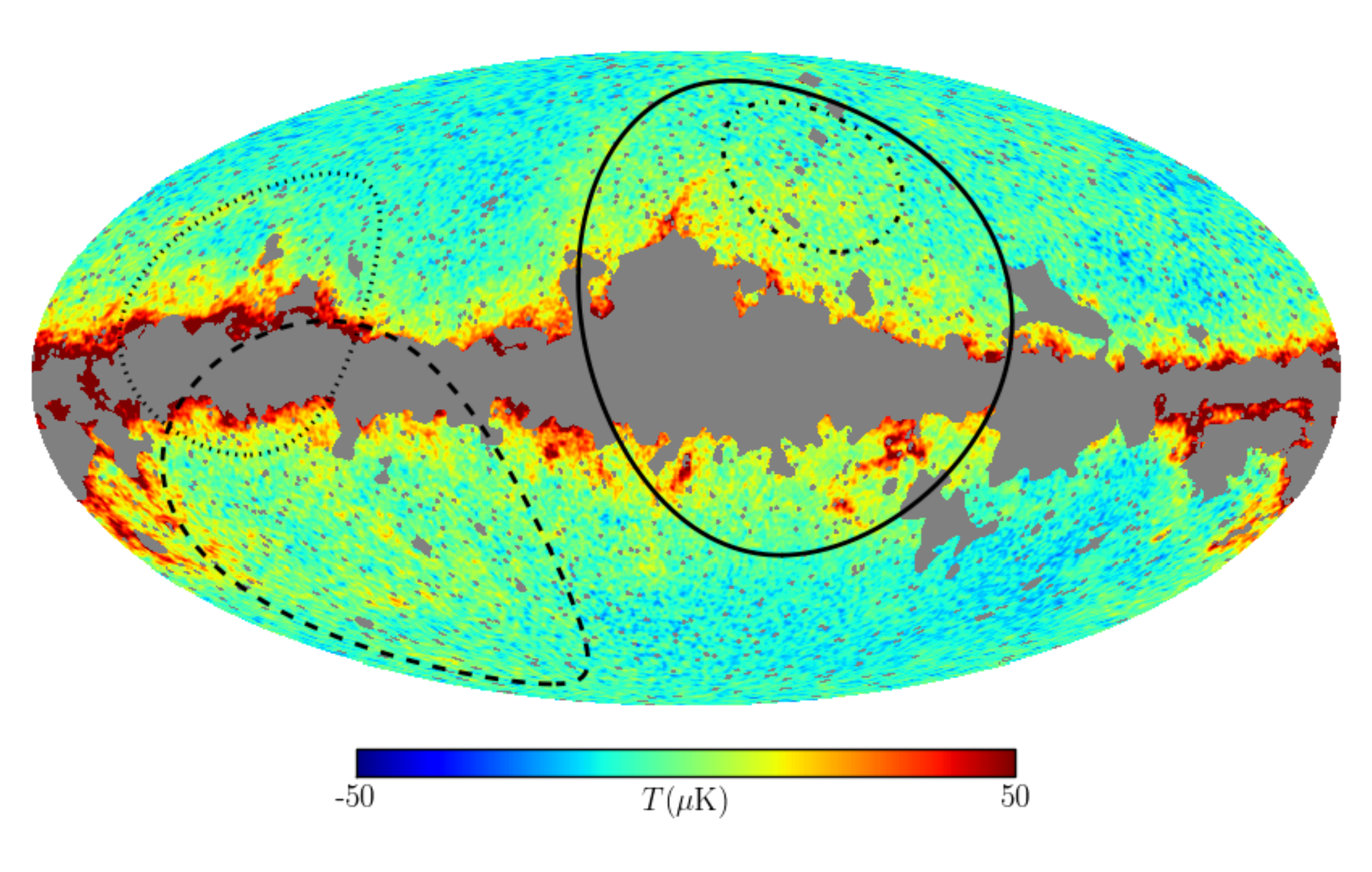}\\
}
\caption{The ILC map generated using $W_j^{\text{rest}}$ (left),
  $W_j^{\text{Loop~I}}$ (middle), and their difference (right), with
  the radio loops superimposed.}
 \label{fig:ilc_diff}
\end{figure*}

We consider a simple model. Suppose that between frequencies
$\nu_{\text{min}}$ and $\nu_{\text{max}}$ (corresponding to bands
$\{j\} = \{1, \mathellipsis N_c\}$ of the $N$ frequency bands) the
area of Loop~I is contaminated by an anomalous black body emission
with temperature $T_s \sim 20 \, \text{K} \gg T_{\text{CMB}}$ and
optical depth $\tau\sim 10^{-6}$, modifying Equation~(\ref{eqn:Sj}) to
\begin{equation}
S_j (\hat{\mathbf n}) = c(\hat{\mathbf n}) +F_j (\hat{\mathbf n}) +
\tau(\hat{\mathbf n}) T_s\Theta(\nu_{\text{min}} \leq \nu_j \leq
\nu_{\text{max}}) \, .
\end{equation}
The error function $\varepsilon(\hat{\mathbf n})$ is then
\begin{eqnarray}
&& \varepsilon(\hat{\mathbf n}) = \sum_{j=1}^N W_j F_j(\hat{\mathbf
    n}) + \sum_{j=1}^{N_c} W_j \tau(\hat{\mathbf n}) T_s \, .
\label{ilc2}
\end{eqnarray}
The second term, will make an anomalous contribution to the
reconstructed CMB signal in the Loop~I region. If this anomaly has the
same frequency dependence as the other foregrounds considered, the ILC
method would have efficiently suppressed its contribution. Conversely,
the existence of the observed anomaly (Section~\ref{sec:peaks}) implies
that, at least locally (in this case in the direction of Loop~I), its
spectrum must be \emph{different}.

We can estimate the spectrum of the anomalous emission from the weight
coefficients $W_j^{\text{Loop~I}}$ derived from a particularly bright
part of Loop~I, and those $W_j^{\text{rest}}$ derived from the CMB sky
(with the 9 yr KQ85 mask) \emph{excluding} the Loop~I region. We adopt
a minimal variance approach (see \cite{Chiang_Nas_Coles,Bennett2012}
for details) and find for the WMAP K, Ka, Q, V, W bands:
$W_j^{\text{rest}}=(-0.181, -0.073, 0.846, 0.460, -0.052)$ and
$W_j^{\text{Loop~I}}=(-0.420, 0.082, 1.654, 0.309, -0.625)$. Since
Loop~I covers a small fraction of the sky, the weight coefficients for
the sky (outside the mask) are similar whether or not the Loop~I
region is included.

However, the weights for the Loop~I region and for the rest of the sky
are \emph{different}. In Figure~\ref{fig:ilc_diff} we show the ILC maps
obtained with both sets of coefficients and their difference map which
clearly indicates the presence of the loops. This means that the
anomaly is efficiently suppressed by the $W_j^{\text{Loop~I}}$. We
have checked which power law indices $\beta$ are ``zeroed'' by these
weights, finding these to be synchrotron-like ($\beta \sim -3$), and
one higher $\beta \sim 1.4$, i.e.,  dust-like but somewhat flatter. We
caution that this is not definitive, e.g., because variation of dust
temperature and temperature conversion factors are ignored.

\section{Physical models of the peaks along Loop~I}
\label{sec:models}

Adopting the synchrotron model \citep{Sarkar1982,Mertsch} for the
radio loops, we have produced maps (with $l \leq 20$) for Loop~I at
WMAP frequencies (22.8-93.5 GHz). With the $W_j^{\text{rest}}$ from
above, the minimum variance method suppresses the contribution from
Loop~I to the ILC map, $T_{\text{ILC,synch}} = \sum_{j=1}^N
W_j^{\text{rest}} T_{\text{synch}} (\nu_j)$, to $2 \, \mu\text{K}$,
compared to $T_{\text{synch}} (22.8 \, \text{GHz})$ of
$160\,\mu\text{K}$. To get a synchrotron contribution to the ILC map
of $20\,\mu$K by extrapolating from 408 MHz, the spectral index would
need to be $-2.4$ (for $W_j^{\text{rest}}$ and $W_j^{\text{loop}}$
respectively) or even as hard as $-2.0$ (for the WMAP region 0
coefficients). We conclude therefore that synchrotron emission
\emph{cannot} be responsible for the anomalies in the WMAP ILC map,
and we should look instead at emission by dust. As noted earlier, the
two emissions are expected to be spatially \emph{correlated} in the
shells of old SNRs like Loop~I.

To further investigate the frequency dependence we have studied the
correlation of the ILC temperature map $T_{\text{ILC}}$ with the WMAP
W- and V-band maps of polarised intensity in temperature units,
$P_{\text{W}}$ and $P_{\text{V}}$. The latter have not entered into
the production of the ILC map and are furthermore expected to be
dominated by polarised dust. Therefore, their correlation with the ILC
map gives us an \emph{independent} handle on the contamination of the
Loop~I region by dust emission. We have computed the ratio of the
polarised intensity in the W- and V-band, averaged over the $\pm
10^{\circ}$ wide ring (see Figure~\ref{fig3}) and find
$\overline{P_{\text{W}}/P_{\text{V}}} = 1.7$ (which changes by less
than $5\%$ with a different ring width e.g., $\pm2^\circ$). If the
temperature anisotropy in this sample is weakly dependent on frequency
$\nu$, and $T(\nu) \approx \overline{T} \simeq 23 \, \mu\text{K}$ (at
least for the WMAP V and W bands) this value of 1.7 indicates the
existence of \emph{two} dust components in the Loop~I region by the
following argument.

Suppose that the anomalous emission is associated with a single
thermal dust component. According to the \textit{Planck} all-sky dust
emission model \citep{PlanckXI,PlanckXVII}, the temperature of the
dust in units of the CMB thermodynamic temperature in the direction
$\mathbf{\hat{n}}$ is
\begin{eqnarray}
\label{equ:powerlaw model}
&&T(\nu,\mathbf{\hat{n}})=T(\nu_0,\mathbf{\hat{n}})
\frac{c(\nu)}{c(\nu_0)}
\frac{\mathrm{e}^{\frac{h\nu_0}{k T_\mathrm{dust}(\mathbf{\hat{n}
        })}}-1}{\mathrm{e}^{\frac{h\nu}{k T_\mathrm{dust}(\mathbf{\hat{n})}}}-1}
\left( \frac{\nu}{\nu_0} \right)^{\beta(\mathbf{\hat{n}})+1}\nonumber
\,, \\ 
&& c(\nu)=\left(2\sinh\frac{x}{2}/x\right)^2,
\end{eqnarray}
where $T_\mathrm{dust}$ is the dust temperature, $x\equiv
h\nu/kT_\mathrm{CMB}\simeq \nu/56.8 \, \text{GHz}$ and $\nu_0$ and
$T(\nu_0,\mathbf{\hat{n}})$ are the reference frequency and
temperature map, respectively. Now, adopting a dust temperature of
$20\,\rm{K}$ \citep{PlanckXI} and assuming that the
polarization fraction $P/T$ is the same in the V- and W-bands (for the
single thermal dust emission model \emph{only}), the spectral index $\beta$
for dust emissivity should be approximately 1. However, inspection of
Fig.~9 of \citet{PlanckXI} shows that the average spectral index for
the Loop~I region (see Figure~\ref{fig3}) is $\beta_\mathrm{pl} \simeq
1.6$. (This number is robust as a different $\beta_\mathrm{pl}$ would
lead to a vastly different extrapolated temperature.) Furthermore, our
finding $\overline{P_{\text{W}}/P_{\text{V}}} = 1.7$ indicates that
the temperature must be closely proportional to frequency, at least
in the V and W bands, i.e., 61--94~GHz.

We thus consider an alternative explanation of the anomalous emission
from Loop~I, viz. \emph{magnetic} dipole emission from dust grains,
arising from thermal fluctuations in the magnetization of grain
materials~\citep{DL_99,2013ApJ...765..159D}. It is likely that this
emission extends beyond the W- and V-bands. It has been noted that
this can be the case if the grain material is ferromagnetic
(i.e., enriched with metallic Fe) or ferrimagnetic (e.g., magnetite,
$\text{Fe}_3\text{O}_4$). Such grains may generate magnetic dipole
emission extending up to 200--300 GHz and it has been argued that this
accounts for the surprisingly strong submillimeter and millimeter emission from the
Small Magellanic Cloud~\citep{2012ApJ...757..103D}. This radiation may
be particularly strong from the galactic loops too, if a higher
fraction of the iron-bearing grains survive after being swept up by
the SN blast wave. A population of Fe grains or grains
with Fe inclusions at a temperature around 20 K would be expected to
produce strong emission due to a ferromagnetic resonance which depends
on shape, going up to $\sim30$~GHz for extreme prolate
grains~\citep{2013ApJ...765..159D}. At 120--130 GHz the magnetic
dipole emission from dust grains will drop below the thermal electric
dipole `vibrational' emission, while well above 200~GHz its
contribution will be negligible relative to thermal emission from
dust.

\section{Conclusion}
\label{sec:conclusion}

We have found evidence of local galactic structures such as Loop~I in
the WMAP ILC map of the CMB which is supposedly fully cleaned of
foreground emissions. This contamination extends to high galactic
latitude so the usual procedure of masking out the Milky Way
\emph{cannot} be fully effective at removing it. It extends to
sufficiently high frequencies that it cannot be synchrotron radiation
but might be magnetic dipole emission from ferro- or
ferrimagnetic dust grains, as suggested by theoretical arguments
\citep{DL_99,2013ApJ...765..159D}. This radiation is expected to be
\emph{polarised} with a frequency dependent polarization fraction.

It has not escaped our attention that as shown in Figure~\ref{fig1}, the
lower part of Loop~I, in particular the new loop S1 identified by
\cite{Wolleben2007}, crosses the very region of the sky from which the
BICEP 2 experiment has recently detected a B-mode polarization signal
at $150 \, \text{GHz}$ \citep{Ade:2014xna}. This has been ascribed to
primordial gravitational waves from inflation because ``available
foreground models'' do not correlate with the BICEP maps. The new
foreground we have identified is however \emph{not} included in these
models. Hence the cosmological significance if any of the detected
B-mode signal needs further investigation. Forthcoming polarization
data from the \textit{Planck} satellite will be crucial in this
regard.


We are indebted to Bruce Draine, Pavel Naselsky, and Andrew Strong for
helpful discussions and to the Discovery Center for support. We thank the 
anonymous referees for critical and constructive comments which helped 
to improve this Letter. We acknowledge use of the \texttt{HEALPix}\footnote{\url{http://healpix.sourceforge.net}}
package~\citep{2005ApJ...622..759G}. Hao Liu is supported by the
National Natural Science Foundation of China (Grant No. 11033003 and
11203024), and the Youth Innovation Promotion Association, CAS.
Philipp Mertsch is supported by DoE contract DE-AC02-76SF00515 and a
KIPAC Kavli Fellowship. Subir Sarkar acknowledges a DNRF Niels Bohr
Professorship.

\bibliography{newbib}

\end{document}